\documentclass[12pt]{iopart}

\usepackage{graphicx}

\usepackage{iopams}

\expandafter\let\csname equation*\endcsname\relax

\expandafter\let\csname endequation*\endcsname\relax

\usepackage{amsmath}
\usepackage[hyphens]{url}
\usepackage{bbm}
\usepackage[]{units}

\usepackage{amssymb}
\usepackage{bm}
\usepackage{hyperref}
\usepackage{braket}
\usepackage{units}
\usepackage{dsfont}

\newcommand{\bne}{\begin{align}}
\newcommand{\ede}{\end{align}}
\newcommand{\bea}{\begin{eqnarray*}}
\newcommand{\eea}{\end{eqnarray*}}
\newcommand{\bnen}{\begin{equation}}
\newcommand{\eden}{\end{equation}}
\newcommand{\bnsn}{\begin{subequations}}
\newcommand{\edsn}{\end{subequations}}
\newcommand{\bean}{\begin{eqnarray}}
\newcommand{\eean}{\end{eqnarray}}
\newcommand{\bna}{\begin{array}}
\newcommand{\eda}{\end{array}}
\newcommand{\st}[1]{_{\text{#1}}}
\newcommand{\bo}[1]{\boldsymbol{#1}}
\newcommand{\E}{\text{e}}

\newcount\colveccount
\newcommand*\colvec[1]{
        \global\colveccount#1
        \begin{pmatrix}
        \colvecnext
}
\def\colvecnext#1{
        #1
        \global\advance\colveccount-1
        \ifnum\colveccount>0
                \\
                \expandafter\colvecnext
        \else
                \end{pmatrix}
        \fi
}

\newcommand{\nfd}{n_{\rm FD}}
\newcommand{\nbe}{n_{\rm BE}}
\newcommand{\Gl}{\Gamma_{\rm L}}
\newcommand{\Gr}{\Gamma_{\rm R}}
\newcommand{\h}{\mathcal{H}}
\newcommand{\gc}{g_{\rm c}}
\newcommand{\nnb}{\nonumber \\ }

\newcommand{\igr}[2][]{\includegraphics[#1]{#2}}

\begin{document}

\title[Valley-resolved spectroscopy in Si TQDs]{Theory of valley-resolved spectroscopy of a Si triple quantum dot coupled to a microwave resonator}
\author{Maximilian Russ}
\address{Department of Physics, University of Konstanz, D-78464 Konstanz, Germany}
\footnote{Both authors contributed equally to this work.}
\author{Csaba G.~P\'{e}terfalvi}
\address{Department of Physics, University of Konstanz, D-78464 Konstanz, Germany}
\footnotemark[1]
\author{Guido Burkard}
\address{Department of Physics, University of Konstanz, D-78464 Konstanz, Germany}
\ead{Guido.Burkard@uni-konstanz.de}

\begin{abstract}
We theoretically study a silicon triple quantum dot (TQD) system coupled to a superconducting microwave resonator. 
The response signal of an injected probe signal can be used to extract information about the level structure by measuring the transmission and phase shift of the output field. This information can further be used to gain knowledge about the valley splittings and valley phases in the individual dots. Since relevant valley states are typically split by several $\mu\text{eV}$, a finite temperature or an applied external bias voltage is required to populate energetically excited states. The theoretical methods in this paper include a capacitor model to fit experimental charging energies, an extended Hubbard model to describe the tunneling dynamics, a rate equation model to find the occupation probabilities, and an input-output model to determine the response signal of the resonator.  

\end{abstract}
\submitto{\JPCM}
\maketitle

Semiconductors with abundant nuclear spin-free isotopes are
increasingly being investigated as host material for spin qubits, e.g.
silicon~\cite{Zwanenburg2013}, germanium\cite{Watzinger2018}, and graphene~\cite{Eich2018,Overweg2018}.
It turns out that most of these materials comprise an electron valley
degree of freedom~\cite{Joyce1993} in the conduction band of the bulk material.
In many nanostructures based on these semiconductor materials,  the
resulting valley splitting is still not fully understood and therefore represents in practice an unpredictable system parameter. It is known that the valley degree of freedom can be described as a pseudo-spin in a two dimensional electron gas (2DEG) whose attributes, i.e., valley-splitting and valley-phase, drastically depend on the interface of the heterostructure~\cite{Friesen2007,Culcer2009,Culcer2010,Veldhorst2015b,Boross2016,Zimmerman2017,Gamble2016,Tariq2019}. A single atomic step can change the quantization axis of the pseudo-spin and the complex phase of the valley-orbit coupling of an electron can be modified by as much as $\pi$~\cite{Zimmerman2017,Gamble2016,Tariq2019}. This has a large impact on silicon quantum computation~\cite{Zwanenburg2013} for most qubit implementations, which use the spin degree of freedom to encode quantum information~\cite{Loss1998}. For multi-qubit quantum processors~\cite{Veldhorst2015,Zajac2017,Watson2018,Yang2019} and multi-spin qubit implementations~\cite{Taylor2013,Russ2017,Sala2018,Russ2018b}, the presence of the valley leads to several non-computational states into which the information can ``leak''. Since the number of leakage states exponentially increases with the number of electrons, the resulting complex energy diagram with a high density of states makes it difficult to find the optimal parameter regimes for encoding and operating such qubits. Therefore, a precise knowledge of the valley structure is required for high-fidelity qubit implementations and operations. 
A lower bound to the valley splitting can be obtained using
ground-state
magnetospectroscopy~\cite{Hada2003,Lim2009,Xiao2010,Lim2011,Borselli2011}. Recent
advances in the coupling of electrons to superconducting microwave
resonators~\cite{Mi2016,Bruhat2016,Samkharadze2016,Stockklauser2017,Mi2017,Mi2018,Samkharadze2017,Landig2017}
allow for precise read-out of the valley splittings in double quantum dots~\cite{Burkard2016,Mi2017}. In this theoretical paper, the technique is extended and adapted to extract the valley splitting and valley phases in a silicon triple quantum dot (TQD) system using such superconducting microwave resonator.

This paper is organized as follows. Firstly, we introduce a general theoretical model of the TQD system in Section~\ref{4sec:theoDes}. For this we use a classical capacitor model to find the electrostatic energies of the electrons and substitute them into an extended Hubbard model to account for hopping of the electrons between the dots (see subsection~\ref{4ssec:Hamiltonian}). Subsequently in subsection~\ref{4ssec:Occupation}, we use a master equation to find the steady state solution of the electron dynamics in the presence of dissipative processes. This allows us to find the corresponding occupation probabilities. Finite temperature effects and an externally applied bias are included in our model.
In subsection~\ref{4ssec:InputOutput}, we consider the response of a superconducting microwave cavity dispersively coupled to the TQD system and use input-output theory to derive analytical expressions for the response signal. Subsequently, in Section~\ref{4sec:resultsSpectr}, we show how one can extract relevant system parameters from the cavity response signal. We explicitly demonstrate the case of a single electron in a triple quantum dot in subsection~\ref{4ssec:single} and the case of three electrons in subsection~\ref{4ssec:three}.

\section{Theoretical description}
\label{4sec:theoDes}

\begin{figure}
\begin{center}
 \igr[width=0.5\columnwidth]{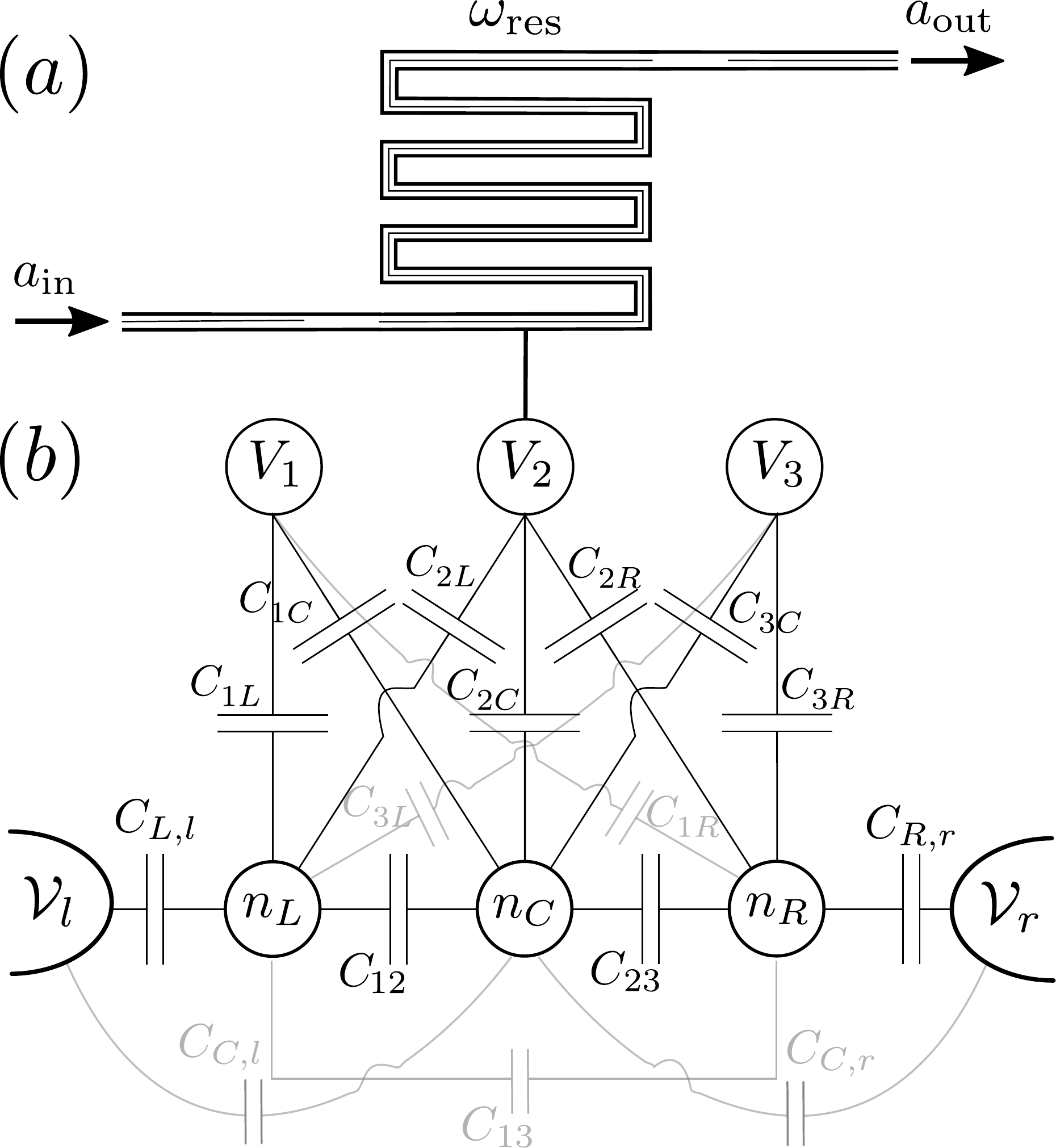}
 \end{center}
  \caption{Illustration of the setup: (a) a superconducting microwave resonator capacitively coupled to a linearly arranged silicon triple quantum dot (TQD) via the center plunger gate $V_{2}$. The cavity is probed with the input signal $a_\text{in}$. Measurement of the transmitted signal $a_{\text{out}}$ can be used to reconstruct the energy landscape of the TQD. (b) Capacitor model of the TQD where each dot is filled with $n_{i}$ electrons ($i=L,C,R$) and capacitively coupled to the electrostatic gates, $V_{1},V_{2},V_{3}$. The $C_{v i}$ denote the capacitance between gate $v=1,2,3$ and dot $i=L,C,R$. The coefficients $C_{i,j}$ describe the capacitances between the electrons in dot $i$ and lead $j=l,r$. Applied voltages to lead reservoirs are denoted by $\mathcal{V}_{j}$. The mutual capacitance $C_{ik}$ describes the electrostatic interaction between the electrons in the QDs~$i$ and~$k$. Black lines capacitively couple neighboring dots, leads or gates, and gray lines denote next-neighbor coupling. Cross-coupling between the electrostatic gates is neglected since it leads only to an overall shift in energy~\cite{Schroeer2007}.}
  \label{4fig:main1}
\end{figure}

We consider a triple quantum dot (TQD) connected to two leads and a superconducting transmission line resonator via the center gate (see Fig.~\ref{4fig:main1}). In order to describe the TQD theoretically, we first construct the bare electron Hamiltonian $\h$ of the system and introduce the interaction to the leads and the microwave resonator later. We consider a basis with 0, 1, 2 or 3 electrons with spin and two-fold valley degeneracy in each dot. For a fixed number of electrons $n_{e}$, there are $\binom{d_{d}\,d_{s}\,d_{v}}{n_{e}}$ possible basis states with $d_{d}\,d_{s}\,d_{v}=3\times2\times2=12$ being the product of the number of dots $d_{d}$, the spin degeneracy $d_{s}$, and the valley degeneracy $d_{v}$. Therefore, we have $N=\sum_{i=0}^{3}\binom{12}{i}=299$ basis states in total. 
Here, we restrict our analysis to the two energetically lowest laying orbital levels. Silicon quantum dots typically have relatively large orbital energies $E\st{orb}$ = $\unit[3-5]{meV}$~\cite{Yang2012,Zajac2016}, thus, the impact of higher orbital levels can be neglected for temperatures $k_{B} T \ll E\st{orb}$ and applied voltage biases $\mathcal{V}_{l}-\mathcal{V}_{r}\ll E\st{orb}/|e|||\alpha||$ where $e$ is the electron charge and $||\alpha||$ is the norm of the lever arm matrix~\cite{Wiel2002}.

\subsection{Hamiltonian}
\label{4ssec:Hamiltonian}

In order to obtain a good agreement of our theoretical studies with experiments we rely on a description with the extended Hubbard model. The electrostatic energies are given by a capacitor model of the TQD, schematically shown in Fig.~\ref{4fig:main1}. The free energy of the triple dot system reads~\cite{Schroeer2007}
\begin{align}
F=\frac{1}{2}\boldsymbol{Q}_{\text{eff}}^{T}\mathcal{C}_{\text{dot}}^{-1}\boldsymbol{Q}_{\text{eff}},
\label{4eq:freeEn}
\end{align}
where $T$ denotes the transposition and
\begin{align}
\boldsymbol{Q}_{\text{eff}}=e\colvec{3}{n_{L}}{n_{C}}{n_{R}}-\mathcal{C}_{\text{gate}}\colvec{3}{V_{1}}{V_{2}}{V_{3}}- \mathcal{C}_{\text{lead}}\colvec{2}{\mathcal{V}_{l}}{\mathcal{V}_{r}}
\end{align}
quantifies the total effective charge on the quantum dots composed of the electron occupation number $n_{i}$ and the applied gate voltages $\bo{V}=(V_{1},V_{2},V_{3})^{T}$. Here $e<0$ denotes the electron charge and $\Delta\mathcal{V}=\mathcal{V}_{l}-\mathcal{V}_{r}$ the applied bias voltages between the left and right leads. The dot capacitance matrix reads
\begin{align}
\mathcal{C}_{\text{dot}}=\left(\begin{array}{ccc}
C_{1} & -C_{12} & -C_{13} \\
-C_{12} & C_{2} & -C_{23} \\
-C_{13} & -C_{23} & C_{3}
\end{array}\right)
\end{align}
which contains the self capacitances $C_{i}$ and the mutual capacitances $C_{12}$, $C_{23}$ and $C_{13}$. The capacitances between the gates and the dots reads
\begin{align}
\mathcal{C}_{\text{gate}}=-\left(\begin{array}{ccc}
C_{1L} & C_{2L} & C_{3L} \\
C_{1C} & C_{2C} & C_{3C} \\
C_{1R} & C_{2R} & C_{3R},
\end{array}\right)
\end{align}
and
\begin{align}
\mathcal{C}_{\text{lead}}=\left(\begin{array}{cc}
-C_{L,l} & 0  \\
-C_{C,l} & -C_{C,r}  \\
0 & -C_{R,r} 
\end{array}\right)
\end{align}
contains the capacitances between the dots and the left and right lead. For later convenience we also define the chemical potential in each dot
\begin{align}
\mu_{L}(\bo{V})=& F\big((n_{L}+1,n_{C},n_{R}),\bo{V}\big)-F\big((n_{L},n_{C},n_{R}),\bo{V}\big)\label{4eq:muL},\\
\mu_{C}(\bo{V})=& F\big((n_{L},n_{C}+1,n_{R}),\bo{V}\big)-F\big((n_{L},n_{C},n_{R}),\bo{V}\big)\label{4eq:muC},\\
\mu_{R}(\bo{V})=& F\big((n_{L},n_{C},n_{R}+1),\bo{V}\big)-F\big((n_{L},n_{C},n_{R}),\bo{V}\big)\label{4eq:muR}.
\end{align}
The total Hamiltonian of the hybrid TQD-resonator system is given by
\begin{align}
\h = H\st{charge} + H\st{Zeeman} +H\st{valley} +H\st{tunnel},
\label{4eq:HamTot}
\end{align}
where the individual contributions are introduced below.

The electrostatic interaction is described by the Hamiltonian
\begin{align}
H\st{charge} =&F\big((0,0,0),\bo{V}\big)\nnb
&+ \sum_{i} \frac{\partial F\big((n_{L},n_{C},n_{R}),\bo{V}\big)}{\partial n_{i}}\bigg|_{n_{L,C,R}=0} \hat{n}_{i} \nnb
&+\frac{1}{2} \sum_{i,j} \frac{\partial^{2} F\big((n_{L},n_{C},n_{R}),\bo{V}\big)}{\partial n_{i}\partial n_{j}}\bigg|_{n_{L,C,R}=0} \hat{n}_{i}\hat{n}_{j}\\
=&F_{0}+e\sum_{i,j}\boldsymbol{V}_{j}\big(\mathcal{C}_{\text{dot}}^{-1}\big)_{ij}\hat{n}_{i}+\frac{e^{2}}{2}\sum_{i,j}\big(\mathcal{C}_{\text{dot}}^{-1}\big)_{ij}\hat{n}_{i}\hat{n}_{j}
\end{align}
with $\hat{n}_{i} = \sum_{s,v}c^{\dagger}_{i,s,v}c_{i,s,v}$ and the free energy $F$ defined in Eq.~\eqref{4eq:freeEn}. Here $c^{\dagger}_{i,s,v}$ ($c^{\dagger}_{i,s,v}$) creates (annihilates) an electron in dot $i=L,C,R$, with spin $s=\uparrow,\downarrow$, and occupying the $v=\pm$ valley state. 

An externally applied magnetic field $\bo{B}$ breaks the spin-degeneracy. Considering a homogeneous magnetic field $\bo{B}=(0,0,B)$, the Zeeman splitting is described by 
\begin{align}
H\st{zeeman} = \frac{E_{Z}}{2}(\hat{n}_{\uparrow}-\hat{n}_{\downarrow})
\end{align}
with $\hat{n}_{s}=\sum_{i,v}c^{\dagger}_{i,s,v}c_{i,s,v}$. The Zeeman energy is $E_{Z}=g\mu_{B}B$ where $g\approx 2$ is the electron g-factor in silicon. To be precise, the electron g-factor depends on the valley and orbital level and is slightly anisotropic giving rise to small effective magnetic field gradients~\cite{Ruskov2018}. Here, this small anisotropy is neglected.

For a silicon heterostructure the two minima in the conduction band~\cite{Zwanenburg2013} give rise to the valley degree of freedom. The valley splitting can be expressed locally in QD~$i$ as~\cite{Friesen2007}
\begin{align}
H_{v,j}=\frac{1}{2}\left(
\begin{array}{cc}
 0 &  \Delta_{j} \\
\Delta_{j}^{*} &  0 
\end{array}
\right)
\end{align}
with the complex quantity $\Delta_{j}=E_{V}^{j}\,\E^{i\phi_{j}}$ consisting of the valley splitting $E_{V}^{j}$ and valley phase $\phi_{j}$ in dot $j=L,C,R$. Because of atomistic defects at the silicon interface the valley pseudo-vector can have a different phase in each dot~\cite{Burkard2016,Zimmerman2017,Tagliaferri2018,Russ2018b}. The valley Hamiltonian in the valley eigenbasis of each dot can be written as
\begin{align}
H\st{valley}=\sum_{i=1}^{3} \frac{E_{V}^{i}}{2}(\hat{n}_{i,+}-\hat{n}_{i,-})
\end{align}
with $\hat{n}_{i,v}=\sum_{s}c^{\dagger}_{i,s,v}c_{i,s,v}$. In this particular choice of representation the valley phase is transferred to the coupling matrix elements between the quantum dots. The single-qubit inter-dot matrix elements in the valley eigenbasis can be expressed as~\cite{Burkard2016}
\begin{align}
c^{\dagger}_{i,s,v}c_{j,s,u} \rightarrow \cos(\theta_{ij})\,c^{\dagger}_{i,s,v}c_{j,s,u} +i \sin(\theta_{ij})\,c^{\dagger}_{i,s,v}c_{j,s,\bar{u}}
\label{4eq:renorm}
\end{align}
with $i,j=L,C,R$, $v=\pm$, $u=\pm$ and $\bar{u}=-u$. The real-valued quantities $\theta_{ij}=(\phi_{i}-\phi_{j})/2$ can be visually interpreted as the angle between the direction of the valley pseudo-spin in dot $i$ and dot $j$. 

Off-diagonal elements of $\h$ allow for coherent hopping of electrons between neighboring quantum dots. In our model hopping is only allowed between basis states with the same total electron number, the same total spin, and conserves the valley. Because of Eq.~\eqref{4eq:renorm}, the tunneling Hamiltonian reads as
\begin{align}
H\st{tunnel} = \sum_{i,j,s,v} t_{ij} &\Big[\cos(\theta_{ij})\, c^{\dagger}_{i,s,v}c_{j,s,v} \nnb
&+i \sin(\theta_{ij})\, c^{\dagger}_{i,s,v}c_{j,s,\bar{v}}\Big],
\label{4eq:HamTun}
\end{align}
with $t_{ij}=t_{ji}$. We define
\begin{align}
t_{L}=&\cos(\theta_{LC})\,t_{12}\label{4eq:tunL},\\
t_{L}^{\prime}=&i\sin(\theta_{LC})\,t_{12}\label{4eq:tunLp},\\
t_{R}=&\cos(\theta_{RC})\,t_{23}\label{4eq:tunR},\\
t_{R}^{\prime}=&-i\sin(\theta_{RC})\,t_{23}\label{4eq:tunRp},\\
t_{13}=&\cos(\theta_{LR})\,t_{13},\\
t_{13}^{\prime}=&-i\sin(\theta_{LR})\,t_{13}.
\end{align}
The tunnel barriers are assumed to be adjusted such that the hopping
matrix elements between the ground states are equal in strength, i.e.,
$|t_{L}|=|t_{R}|$. Note that, to warrant a unique stationary solution (see below), we chose the valley phases $\theta_{12}\neq n_{1}\,\frac{\pi}{2}$ and $\theta_{23}\neq n_{2}\,\frac{\pi}{2}$ with integer $n_{1},n_{2}$. Because of the linear alinement of the TQD direct hopping between the left and the right dot becomes negligible, thus, we set $t_{13}=0$. As a consequence the valley phase $\theta_{LR}$ becomes undetectable.

\subsection{Occupation probabilities}
\label{4ssec:Occupation}

In order to calculate the occupation probabilities of the dots in the stationary state, we assume that incoherent transitions can occur between the eigenstates of $\h$, both internally and via electron hopping between the TQD and the leads. These incoherent interactions with the environment can be taken into account with the Lindblad master equation
\bnen
\dot{\rho}=-\frac{i}{\hbar}[\h,\rho]+\mathcal{D}(\rho),\label{4eq:Lindblad}
\eden
where $\hbar$ is the reduced Planck constant and $\rho$ is the density matrix. The dissipative part $\mathcal{D}(\rho)$ consists of the following terms
\begin{align}
\mathcal{D}(\rho) =
                      \sum_{\substack{v=\pm,s=\uparrow\downarrow\\i=L,R}}
  \Gamma_{i} \Big(  D\big[c_{i,s,v}^{\dagger}\big](\rho)  
 +\,\, D\big[c_{i,s,v}\big](\rho) \Big) 
  + \sum_{\lambda,\lambda^{\prime}}\,\,\,\,\,\,
                                                            \tau^{-1}_{\lambda\lambda^{\prime}}  D\big[b_{\lambda\lambda^{\prime}}\big](\rho), \label{4eq:lb}
\end{align}
Here $D\big[O \big](\rho)=O^{\dagger}\rho O-(\rho OO^{\dagger}+
OO^{\dagger}\rho)/2$ is the usual Lindbald super operator,
$\Gamma_{i}$ is the transition rate from the lead $i=L,R$ to the dot
$i$, and the operators $c_{i,s,v}^\dagger$, and $c_{i,s,v}$ create and
annihilate an electron in dot $i$ and valley $v$ with spin $s$,
respectively. The first and second terms of \eqref{4eq:lb} correspond to the flow of
electrons from lead $i=l,r$ into dot $i=L,R$ and in the opposite
direction, out of the dot to the lead. The third term in \eqref{4eq:lb} describes excitations within the TQD, i.e. incoherent interactions with a bosonic bath, such as phonons, that can induce transitions from one eigenstate of $\h$ to the other with the same total number of electrons in the dots with the same total spin. The operator $b_{\lambda\lambda^{\prime}}=\left| \lambda\right\rangle \left\langle \lambda^{\prime}\right|$ takes the system from an initial state $\left| \lambda^{\prime}\right\rangle$ to a final state $\left| \lambda\right\rangle$ with a transition rate $\tau^{-1}_{\lambda\lambda^{\prime}}$.

We assume that the level broadenings, caused by the interaction with the leads and the bosonic bath, are smaller than the level splittings between the eigenstates of $\h$ which we ensure by an external magnetic field $B$. This is the so-called secular approximation~\cite{Breuer2007}, which results in a steady-state density matrix $\rho$ diagonal in the eigenbasis of $\h$. 
This significantly simplifies the Lindblad equation~\eqref{4eq:Lindblad}, where the commutator vanishes and after taking the matrix representation of the remaining dissipative term in the eigenbasis of $\h$, we obtain Redfield equations for the diagonal elements of the steady-state solution
\begin{align}
0=\dot{\rho}_{n}=\sum_{\substack{m\\j=L,R}}   \Gamma_{j} \left(
  \rho_{m} c_{mjn} - \rho_{n} c_{njm}   
  +\, \rho_{m} c_{njm} - \rho_{n} c_{mjn} \right) 
 + \sum_{m\neq n}  \left(\tau^{-1}_{nm} \rho_{m} - \tau^{-1}_{mn} \rho_{n}\right),
  \label{4eq:dlb}
\end{align}
where $m,n$ runs over all diagonal elements of $\rho$.
The terms in \eqref{4eq:dlb} are approximations of their respective
counterparts in \eqref{4eq:lb}.
Here $\rho_{n}\equiv\bra{n}\rho\ket{n}$  is the $n$-th diagonal element of the density matrix $\rho$, and $c_{mjn}=\sum_{v,s} \left|\bra{m} c_{j,s,v} \ket{n} \right|^2$, which can be finite only if there is one more electron in state $\ket{n}$ than in $\ket{m}$. 

We can extend this description toward finite temperatures in the leads with the following replacement rules in Eq.~\eqref{4eq:dlb}
\begin{subequations}
\bean
\rho_{m} c_{mjn} && \rightarrow \rho_{m} c_{mjn} n_{njm}\\
\rho_{n} c_{mjn} && \rightarrow \rho_{n} c_{mjn} n_{mjn},
\eean
\end{subequations}
where $n_{mjn} =\nfd\big(E_{m}-E_{n}+(\nu_{m}-\nu_{n})|e|\mathcal{V}_{j},T\big)$, $E_{m}$ and $E_{n}$ are the eigenenergies, $\nu_{m}$ and $\nu_{n}$ the number of electrons in the given eigenstates of $\h$, and
\bnen
\nfd(\delta E_{j},T)=\frac{1}{\exp(\delta E_{j} /(k_{\rm B}T))+1}
\label{4eq:fd}
\eden
is the Fermi-Dirac distribution of the electrons in the lead $j$, with $k_{\rm B}$ and $T$ being the Boltzmann constant and the electron temperature. 

To include finite temperature effects in Eq.~\eqref{4eq:dlb} the $\tau^{-1}_{mn}$ transition rates in (\ref{4eq:dlb}) are redefined as $\tau^{-1}_{mn}=\gamma_{mn}\, \nbe(E_m-E_n,T)$ with the temperature dependent prefactor
\bnen
\nbe(\delta\! E,T)=\frac{1}{\exp(\left|\delta\! E \right|/(k_{\rm B}T))-1}+\Theta(-\delta\! E)
\eden
which accounts for the Bose-Einstein statistics of the environmental thermal bath, that is assumed to be in thermal equilibrium with the electronic system and having an approximately constant density of states in the relevant energy window of the transitions. Moreover, $\Theta(\cdot)$ denotes the Heaviside step function. 

We use the following phenomenological model to describe incoherent decay from $\ket{m}\rightarrow\ket{n}$ with rate
\begin{align}
\gamma_{mn}=\bra{m}^{|\cdot|} \,\,\Gamma\st{eff}\,\,\ket{n}^{|\cdot|},
\label{4eq:gamma}
\end{align}
where $\ket{n}$ denotes the eigenstate of the unperturbed system given in Eq.~\eqref{4eq:HamTot} with eigenenergy $E_{n}$. 
In our model $\ket{m}^{|\cdot|}$ denotes the absolute-valued
eigenvector obtained by taking the absolute value of each entry in
$\ket{m}$
in the eigenbasis of ${\cal H}$ and the matrix elements of the effective decay rate read
\begin{subequations}
\begin{align}
\Gamma_{\text{eff},pq}=\,\,&
      \gamma_{c}  \bra{p}\sum\limits_{\substack{s,v\\|i-j|=1}}c^{\dagger}_{i,s,v}c_{j,s,v}\ket{q} \\
      +&\gamma_{v}  \bra{p}\sum\limits_{\substack{s,v\\|i-j|\leq1}}c^{\dagger}_{i,s,v}c_{j,s,\overline{v}}\ket{q} \\
      +&\gamma_{s}  \bra{p}\sum\limits_{\substack{s,v,v^{\prime}\\|i-j|\leq1}}c^{\dagger}_{i,s,v}c_{j,\overline{s},v^{\prime}}\ket{q},
\end{align}
 \end{subequations}
with $i,j\in{L,C,R}$, $v=\pm$, $s=\uparrow,\downarrow$, and $\overline{v}$ ($\overline{s}$) being the flipped valley (spin).
Here, $\gamma_{c}$ denotes the pure charge relaxation rate and $\gamma_{v}$ describes the relaxation rate involving a valley flip. We neglect any spin-related decays, $\gamma_{s}=0$ due to the long spin-flip time on the order of milliseconds. 
Because of $\gamma_{c}\gg\gamma_{v}$ a decay channel, where both the charge and the valley changes, is always limited by the smaller decay rate and the valley decay $\gamma_{v}$ serves as a bottleneck of the process. 
The matrix elements are between eigenstates $\ket{p},\ket{q}$ of $H_{0}=H\st{charge} + H\st{Zeeman} +H\st{valley}$.

Eq.~(\ref{4eq:dlb}) can cast into a more concise form, which also reflects the temperature dependence
\begin{align}
0=\dot{\rho}_n=\sum_{m\neq n} \left( \Gamma_{nm} \rho_{m} - \Gamma_{mn} \rho_{n} \right) \nnb + \sum_{m\neq n} \left( \tau^{-1}_{nm} \rho_{m} - \tau^{-1}_{mn} \rho_{n} \right),
\label{4eq:master_concise}
\end{align}
where the total decay rate $\Gamma_{mn}$ of the state $\ket{n}$ to state $\ket{m}$ with one electron hopping on or off the TQD is given by
\begin{align}
\Gamma_{mn}=\sum_{j=L,R} \Gamma_{j} \left(c_{mjn} + c_{njm}\right)n_{mjn}.
\end{align}
Note that depending on the direction of the hopping, either $c_{mjn}$ or $c_{njm}$ will be zero. This set of classical rate equations can also be formulated as a matrix equation $M \bo{\rho}=0$, where $\bo{\rho}$ is a vector of the diagonal elements of the density matrix $\rho$. The steady-state solution in the secular approximation is thus provided by the nullspace of the matrix $M$, as a normalized vector of the probabilities $\overline{\rho}_k$ for finding the system in its $k$th eigenstate. If the calculation of the nullspace of $M$ does not return the expected, physically meaningful result because of numerical inaccuracies, then the direct integration of Eq.~\eqref{4eq:master_concise} with the initial condition of a thermal distribution can deliver the correct solution.

\subsection{Input-ouput theory}
\label{4ssec:InputOutput}

For read-out of the energies in the system, one can directly connect the oscillating voltage generated inside the microwave resonator to one of the gate potentials [see Fig.~\ref{4fig:main1}~(a)]. The response of the system to a microwave probe field due to this electric dipole coupling can be determined using cavity input-output theory~\cite{Collett1984}. We assume that the microwave field can induce transitions between all energy levels of the TQD, whereby transitions between neighboring energy levels are more likely for low temperatures and bias voltages. Following the calculation given in Refs.~\cite{Kulkarni2014,Burkard2016,Benito2017,Kohler2018} the transmission coefficient $A$ of the output signal for the TQD is
\bnen
A=\frac{-i\sqrt{\kappa_1 \kappa_2}}{\omega\st{res}-\omega_{P}-i\kappa/2+\gc \sum_{m=1}^{N} \sum_{n=1}^{N} d_{m,n} \chi_{m,n}}.\label{4eq:A}
\eden
The electric susceptibility of the TQD is given by
\bnen
\chi_{n,m}=\frac{-\gc d_{m,n}(\overline{\rho}_n-\overline{\rho}_{m})}{E_{m}-E_n-\omega_{P}-i(\gamma^{*}_{mn}+\tau^{-1}_{mn}/2)}.
\eden
Here $d_{n,m}$ is the dipole matrix element pertaining to the $n
\rightarrow m$ transition, $\tau^{-1}_{mn}$ is the relaxation rate
[see Eq.~\eqref{4eq:gamma}], and
$\gamma^{*}_{mn}=\gamma\st{dep}\sum_{i}(\partial(E_{m}-E_n)/ \partial
V_{i})/||\alpha||$ describes pure dephasing with rate $\gamma\st{dep}$
due to charge noise~\cite{Russ2017}. The total cavity decay rate is
$\kappa=\kappa_1+\kappa_2+\kappa_i$, where $\kappa_1$ and $\kappa_2$
are the photon decay rates through the input and output ports and
$\kappa_i$ is the intrinsic photon decay rate. The probe frequency and
the cavity resonance frequency are denoted by $\omega_{P}$ and
$\omega\st{res}$, and $\gc$ (also commonly known as $\gc=2g_0$) is the
electric dipole coupling strength. The charge noise is coupled through
the electrodes to the electrons via the lever arm matrix $\alpha=e\,\mathcal{C}_{\text{dot}}^{-1}\mathcal{C}_{\text{gate}}$.
The summation in Eq.~\eqref{4eq:A} runs over all the possible transitions within the $N$-electron states, and the eigenstates are indexed with increasing eigenenergies $E_n$. The dipole matrix can be calculated easily in the basis of $\h$, by taking the derivative
\begin{align}
\mathcal{D}=\frac{\partial\h(V_{1},V_{2},V_{3})}{\partial V_{2}},
\end{align}
with gate $V_{2}$ connected to the resonator [see Fig.~\ref{4fig:main1}~(a)]. The dipole matrix elements are then accordingly defined as 
\begin{align}
d_{m,n}=\bra{m}\mathcal{D}\ket{n}.
\end{align}

\section{Results}
\label{4sec:resultsSpectr}
Our goal is to extract information about the energetic structure, in particular the valley splitting and valley phase, of the triple quantum dot system from a measurement of the output signal of the microwave resonator. We expect, in analogy to Ref.~\cite{Burkard2016}, that the finite dipole moment at avoided crossings in the triple quantum dot system yields measurable features in the output signal. Ideally, the location of these features as a function of detuning parameters allows us to reconstruct the energy spectrum of the triple dot. In order to limit the number of anti-crossings we first analyze the case of a single electron in the triple dot. Afterwards, we use the collected information to interpret the case of three electrons which has broad interest due to the realization as an exchange-only qubit~\cite{Russ2017}.

We further consider a homogeneous magnetic field with Zeeman spin splitting $E_{Z}^{i}=E_{Z}=\unit[0.3]{meV}$ (corresponding to $\approx\unit[2.6]{T}$ in silicon) larger than typical valley splittings $E_{Z}>E_{V}^{i}$ for $i=L,C,R$ in SiGe quantum dots. The presence of a magnetic field allows us to ignore the spin degree of freedom and focus solely on valley physics. The remaining simulation parameters are listed in appendix~\ref{4ch:app}.

\begin{center}
\begin{figure}
\centering\igr[width=0.8\textwidth]{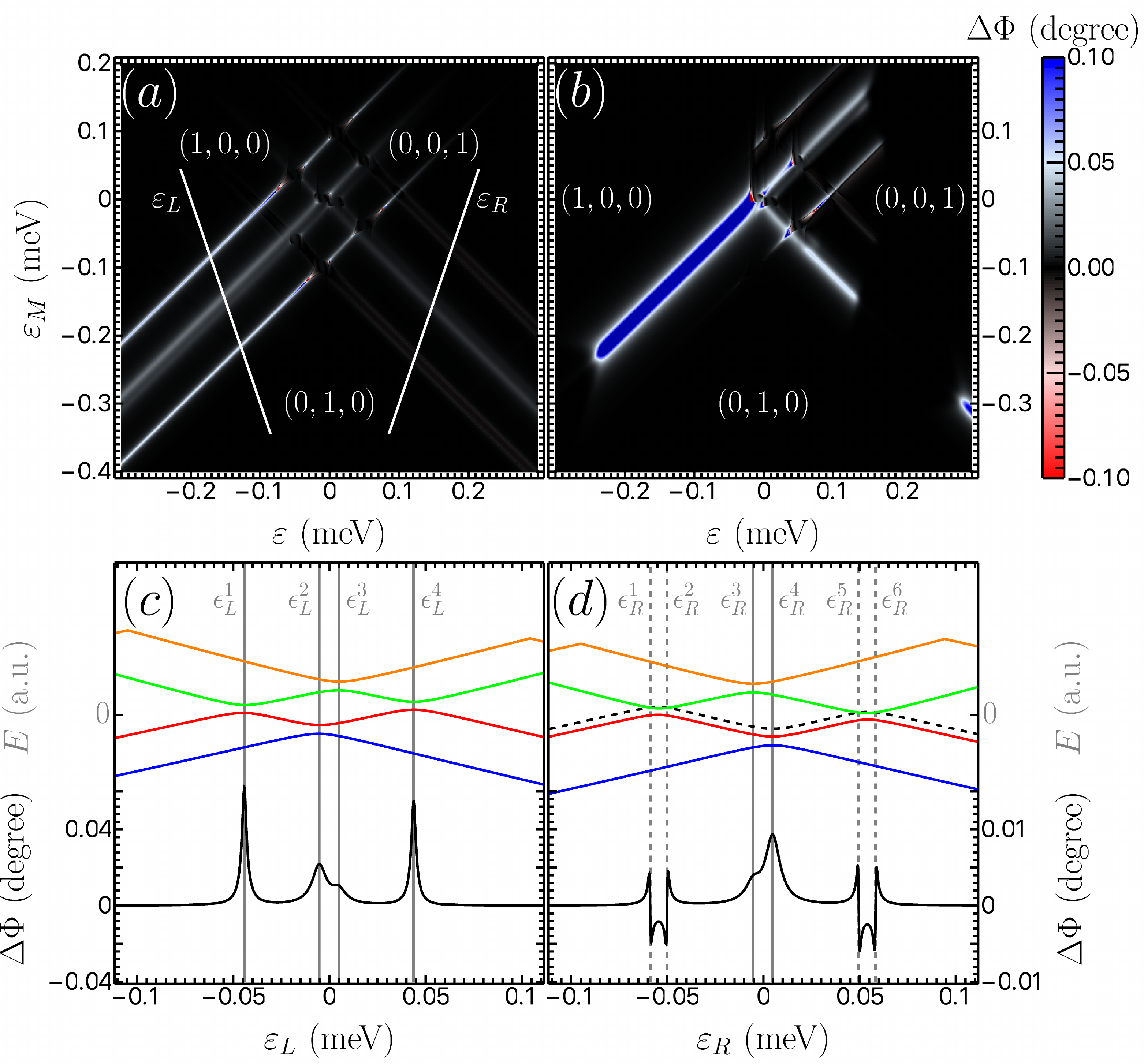}
\caption{(a) Calculated phase response $\Delta\Phi=\text{arg}(A)$ of the probe signal of the TQD filled with a single electron coupled to the microwave cavity in thermal equilibrium at $T=\unit[1]{K}$ and without applied voltage bias $\Delta\mathcal{V}=0$ as a function of the TQD detuning parameters $\varepsilon$ and $\varepsilon_{M}$. Here $(n_{L},n_{C},n_{R})$ denotes the occupation of dot $i$ by $n_{i}$ with $i=L,C,R$ electrons. The left and right white lines are cuts along the double quantum dot (DQD) detuning parameters $\varepsilon_{L}$ and $\varepsilon_{R}$ while keeping $\varepsilon_{g}$ and the respective orthogonal detuning parameter fixed. This allows for the investigation of the (1,0,0)-(0,1,0) and (0,0,1)-(0,1,0) charge transition. (b) Phase response of the probe signal of the TQD coupled to the microwave cavity for $T=\unit[30]{mK}$ and applied bias $\Delta\mathcal{V}=\unit[0.3]{mV}$. Similar features are visible as for the high-temperature phase response with zero bias. 
 (c) Cut along $\varepsilon_{L}$ (white line in (a)) and the energy of the four lowest eigenstates, $E_{1}$, $E_{2}$, $E_{3}$, $E_{4}$, is plotted as function of $\varepsilon_{L}$. The peaks $\epsilon_{L}^{1}$,  $\epsilon_{L}^{2}$,  $\epsilon_{L}^{3}$, and $\epsilon_{L}^{4}$ in the phase response correspond to an anti-crossing between states $\ket{E_{2}}\leftrightarrow \ket{E_{3}}$, $\ket{E_{1}}\leftrightarrow \ket{E_{2}}$, $\ket{E_{3}}\leftrightarrow \ket{E_{4}}$, and $\ket{E_{2}}\leftrightarrow \ket{E_{3}}$.
 (d) Cut along $\varepsilon_{R}$ [white line in (a)]. For identification of the avoided crossings also the energy of the four lowest eigenstates, $E_{1}$, $E_{2}$, $E_{3}$, $E_{4}$, is plotted as function of $\varepsilon_{R}$. For convenience the $E_{2}+\hbar\omega\st{res}$ (black-dashed) is also shown. The peaks $\epsilon_{R}^{3}$ and $\epsilon_{R}^{4}$ in the phase response correspond to an anti-crossing between states $\ket{E_{1}}\leftrightarrow \ket{E_{2}}$ and $\ket{E_{3}}\leftrightarrow \ket{E_{4}}$. The very sharp peaks $\epsilon_{R}^{1}$ and $\epsilon_{R}^{2}$ and $\epsilon_{R}^{5}$ and $\epsilon_{R}^{6}$ correspond to the condition $E_{3}-E_{2}=\hbar\omega\st{res}$ (crossings of black-dashed line). The void area between the (0,1,0) and (0,0,1) originated from a steady state with a completely depleted triple quantum dot.}
  \label{4fig:main2}
\end{figure}
\end{center}

\subsection{One electron in a triple quantum dot}
\label{4ssec:single}

Considering a single electron in the TQD the total Hamiltonian reads
\begin{align}
\h_{1e}=H\st{charge,1e}+H\st{valley,1e}+H\st{Zeeman,1e}+H\st{tunnel,1e}
\label{4eq:Ham1e}
\end{align}
which can be obtained from Hamiltonian~\eqref{4eq:HamTot} using $N=1$. The charge Hamiltonian (here in the basis $\ket{L},\ket{C},\ket{R}$) 
\begin{align}
H\st{charge,1e}=\left(
\begin{matrix}
\varepsilon-\frac{\varepsilon_{M}}{3} & 0 & 0\\
0 & \frac{2\varepsilon_{M}}{3} & 0\\
0& 0 &-\varepsilon-\frac{\varepsilon_{M}}{3}
\end{matrix}\right)+\frac{\varepsilon_{g}}{2}\mathds{1}_{3}
\end{align}
contains the electrostatic interactions from the capacitor model.
The two detuning parameters are defined as
\begin{align}
\varepsilon= &(\mu_{L}-\mu_{R})/2\label{4eq:det1ed}\\
\varepsilon_{M}=&\mu_{C}-(\mu_{L}+\mu_{R})/2
\label{4eq:det1em}
\end{align}
where $\mu_{i}$ is the chemical potentials of quantum dot $i=L,C,R$ given in Eqs.~\eqref{4eq:muL}-\eqref{4eq:muR} with $\mathcal{N}=(0,0,0)$. The average energy in the TQD is then given by
\begin{align}
\varepsilon_{g}=(\mu_{C}+\mu_{L}+\mu_{R})/3.
\end{align}
Through a variation of $\varepsilon_{g}$ the total amount of electrons inside the TQD can be adjusted.
Furthermore, we introduce two additional detuning parameters
\begin{align}
\varepsilon_{L} =& (\mu_{L}-\mu_{C})/2,\\
\varepsilon_{R} =& (\mu_{R}-\mu_{C})/2.
\end{align}
These two detuning parameters have two implications.
Firstly, $\varepsilon_{L}$ and $\varepsilon_{R}$ allow for a simple investigation of the (1,0,0)-(0,1,0) and (0,0,1)-(0,1,0) charge transitions. At these transitions the TQD behaves like a DQD with one charge state highly separated in energy. This effectively reduces the system to a conventional charge qubit.
Secondly, unlike the set, $\varepsilon$, $\varepsilon_{M}$, and $\varepsilon_{g}$, the set $\varepsilon_{L}$, $\varepsilon_{R}$, and $\varepsilon_{g}$ does not form an orthogonal set. Therefore, it is impossible to sweep through the left charge qubit along $\varepsilon_{L}$ while keeping the average energy $\varepsilon_{g}$ and the right-center detuning $\varepsilon_{R}$ constant. Respective cuts along $\varepsilon_{L}$ and $\varepsilon_{R}$ seem to be non-orthogonal to the respective charge transition in $(\varepsilon,\varepsilon_{M})$ space [see Fig.~\ref{4fig:main2}~(a)].

\subsubsection{Zero bias}
The valley degeneracy effectively creates two copies of the charge states which are coupled by the valley non-conserving tunnel amplitudes. Therefore, instead of a single anti-crossing between charge states we expect to see four anti-crossings~\cite{Burkard2016}. In Fig.~\ref{4fig:main2}~(a) the phase shift of the cavity signal for the single electron is shown as a function of the two detuning parameters $\varepsilon,\varepsilon_{M}$. At the (1,0,0)-(0,1,0) and the (0,0,1)-(0,1,0) charge transitions we see the splitting of a single line into three and five distinct lines. 
A cut along the left-center detuning $\varepsilon_{L}$ shows in comparison to the level diagram that the phase responses directly match the corresponding valley splittings [see Fig.~\ref{4fig:main2}~(c)]. We observe a phase response (peak) at $\epsilon^{1}_{L}=-\left(E_{V}^{L}+E_{V}^{C}\right)/4$, $\epsilon^{2}_{L}=\left(E_{V}^{L}-E_{V}^{C}\right)/4$, $\epsilon^{3}_{L}=\left(E_{V}^{L}-E_{V}^{C}\right)/4$, and $\epsilon^{4}_{L}=\left(E_{V}^{L}+E_{V}^{C}\right)/4$. 

A cut along the right-center detuning $\epsilon_{R}$ shows a very similar phase response [see Fig.~\ref{4fig:main2}~(d)]. We observe a phase response (peak) at $\epsilon^{3}_{R}=-\left(E_{V}^{R}-E_{V}^{C}\right)/4$ and $\epsilon^{4}_{R}=\left(E_{V}^{R}-E_{V}^{C}\right)/4$. However, there is no phase response at $\varepsilon_{R}=\mp\left(E_{V}^{R}+E_{V}^{C}\right)/4$ but instead two phase responses (each a sharp dip followed by a sharp peak) at $\epsilon^{1}_{R}=\unit[-58.8]{\mu eV}$, $\epsilon^{2}_{R}=\unit[-50.1]{\mu eV}$ and $\epsilon^{5}_{R}=\unit[49.6]{\mu eV}$, $\epsilon^{6}_{R}=\unit[58.3]{\mu eV}$ (simulation parameters are listed in appendix~\ref{4ch:app}). This splitting into two signals appears if the energy splitting at the avoided crossing is smaller than the resonator frequency, $2t_{R}^{\prime}<\hbar\omega\st{res}$. The microwave resonator becomes resonant with the ground-state excited-state transition $\hbar\omega\st{res}=E_{3}-E_{2}$ at exactly two points [see crossing between dashed and solid lines in Fig.~\ref{4fig:main2}~(d)]. For small tunnel couplings $|t_{R}^{\prime}|\ll|E_{V}^{L}+E_{V}^{C}|/4$ the left anti-crossing between the first and second excited state in Fig.~\ref{4fig:main2}~(d) can be approximated by an isolated two-level system with energy splitting 
\begin{align}
\Delta E_{R,1}=2\sqrt{(\varepsilon_{R}-\tilde{\epsilon}_{R}^{1})^{2}+|t_{R}^{\prime}|^{2}},
\label{4eq:splittLeft}
\end{align}
where $\tilde{\epsilon}^{1}_{R}$ is the position in $\varepsilon_{R}$ detuning space. From the equation above it follows that $(\epsilon^{1}_{R}+\epsilon^{2}_{R})/2=-\left(E_{V}^{R}+E_{V}^{C}\right)/4$. Similarly, we find the position of the right anti-crossing between the first and second excited state at $(\epsilon^{5}_{R}+\epsilon^{6}_{R})/2=\left(E_{V}^{R}+E_{V}^{C}\right)/4$. 

In total we extract the valley splittings $E_{V}^{L}=\unit[77.7]{\mu eV}$, $E_{V}^{C}=\unit[98.2]{\mu eV}$, and $E_{V}^{R}=\unit[118.6]{\mu eV}$ which are roughly $2\%$ smaller than the input settings $\widetilde{E}_{V}^{L}=\unit[80]{\mu eV}$, $\widetilde{E}_{V}^{C}=\unit[100]{\mu eV}$, and $\widetilde{E}_{V}^{R}=\unit[120]{\mu eV}$. We attribute this small systematic error to a deformation of the energy levels due to the mixing of the different levels via tunneling. To mitigate these kind of errors the cuts along $\varepsilon_{L}$ and $\varepsilon_{R}$ can be performed further away from the triple point, $\varepsilon=\varepsilon_{M}=0$ where all three charge configurations intersect. 
Note that we assumed an electron temperature $T=\unit[1]{K}$ to occupy the excited states and see features of the excited valley states in Fig.~\ref{4fig:main2}~(a). The phase response of the cold simulation with $T=\unit[30]{mK}$ but applied bias $\Delta\mathcal{V}=\unit[0.3]{meV}$ between the two leads shows similar features [see Fig.~\ref{4fig:main2}~(b)] in the vicinity of the triple point. However, there is only a small energy window in which a finite charge current is possible~\cite{Schroeer2007} and higher lying valley states have a finite occupation probability. At the relevant (1,0,0)-(0,1,0) and the (0,0,1)-(0,1,0) charge transitions the charge current is blocked suppressing any probe signal from higher states (see appendix~\ref{4ssec:current}). An alternative measurement scheme for small temperature is discussed in the next subsection.

The extraction of the valley phase is a more challenging task. Following the procedure in Ref.~\cite{Mi2017} the valley phase can be estimated by fitting the amplitudes of the phase response for the avoided crossings at $\epsilon^{1}_{L}$, $\epsilon^{2}_{L}$, $\epsilon^{3}_{L}$, and $\epsilon^{4}_{L}$ in Fig.~\ref{4fig:main2}~(c) to the tunnel couplings $t_{L}$ and $t_{L}^{\prime}$.  Unfortunately, the fitting includes two more unknown parameters (taking into account charge noise) making the fits hard and unstable. Furthermore, this method requires large tunnel couplings $2|t_{j},t^{\prime}_{j}|>\hbar\omega\st{res}$ with $j=L,R$ to gain a single response signal which for our parameter setting is not fulfilled for the (0,1,0)-(0,0,1) charge transition. Then the tunnel coupling strength $|t_{R}|$ and $|t_{R}^{\prime}|$ can be extracted by fitting to the energy gap. For small tunnel amplitudes Eq.~\eqref{4eq:splittLeft} provides a sufficient approximation. Alternatively, for a frequency tunable resonator~\cite{Landig2017} the tunnel couplings can be extracted using spectroscopy by observing the splitting of the signal into two signals mentioned above. Using Eqs.~\eqref{4eq:tunL}-\eqref{4eq:tunRp} the two valley phases are given by $\theta_{LC}=\tan^{-1}|t_{L}^{\prime}/t_{L}|=\unit[0.23]{\pi}$ and $\theta_{RC}=\tan^{-1}|t_{R}^{\prime}/t_{R}|=\unit[-0.2]{\pi}$. 

The methods introduced here do not provide a way to measure the valley angle between the left and right valley pseudo-spin $\theta_{LR}$. In our simplified picture for the tunneling between the dots, a direct tunnel matrix element $t_{13}$ between the left and right dot is set to zero which is close to the real scenario for a linear array. For a triangular geometry of the triple dot all tunnel matrix elements are finite and the remaining valley phase difference $\theta_{LR}$ can be directly measured by performing the same type of measurement to extract $\theta_{LC}$ and $\theta_{RC}$ at the (1,0,0)-(0,0,1) charge transition. 
This requires the comparison of the tunnel couplings $t_{13}$ and $t_{13}^{\prime}$. Furthermore, we note that in the presence of a triangular geometry a closed path can give rise to a non-vanishing geometric phase. This can in principle also be used to probe the valley in a complementary way.

{\renewcommand{\arraystretch}{1.5}
\begin{table}
\begin{center}
\begin{tabular}{|c||c|c|c|}
\hline\hline
& $\varepsilon$ & $\varepsilon_{M}$ & $E\st{ex}$\\
\hline
 $\epsilon_{Q}^{1}$ &  $\frac{E_{V}^{L}}{4}-\frac{E_{V}^{R}}{4}$ & $\frac{E_{V}^{C}}{2}-\frac{E_{V}^{L}}{4}-\frac{E_{V}^{R}}{4}$ & $0$ 
 \\
 $\epsilon_{Q}^{2}$ &  $-\frac{E_{V}^{L}}{4}-\frac{E_{V}^{R}}{4}$ & $\frac{E_{V}^{C}}{2}+\frac{E_{V}^{L}}{4}-\frac{E_{V}^{R}}{4}$ & $E_{V}^{L}$ \\
 $\epsilon_{Q}^{3}$ &  $\frac{E_{V}^{L}}{4}-\frac{E_{V}^{R}}{4}$ & $-\frac{E_{V}^{C}}{2}-\frac{E_{V}^{L}}{4}-\frac{E_{V}^{R}}{4}$ & $E_{V}^{C}$ \\
 $\epsilon_{Q}^{4}$ &  $\frac{E_{V}^{L}}{4}+\frac{E_{V}^{R}}{4}$ & $\frac{E_{V}^{C}}{2}-\frac{E_{V}^{L}}{4}+\frac{E_{V}^{R}}{4}$ & $E_{V}^{R}$ \\
 $\epsilon_{Q}^{5}$ & $-\frac{E_{V}^{L}}{4}-\frac{E_{V}^{R}}{4}$ & $-\frac{E_{V}^{C}}{2}+\frac{E_{V}^{L}}{4}-\frac{E_{V}^{R}}{4}$ & max$(E_{V}^{L},E_{V}^{C})$ 
 \\
 $\epsilon_{Q}^{6}$ &  $\frac{E_{V}^{R}}{4}-\frac{E_{V}^{L}}{4}$ & $\frac{E_{V}^{C}}{2}+\frac{E_{V}^{L}}{4}+\frac{E_{V}^{R}}{4}$ & max$(E_{V}^{L},E_{V}^{R})$ 
 \\
 $\epsilon_{Q}^{7}$ &  $\frac{E_{V}^{L}}{4}+\frac{E_{V}^{R}}{4}$ & $-\frac{E_{V}^{C}}{2}-\frac{E_{V}^{L}}{4}+\frac{E_{V}^{R}}{4}$ & max$(E_{V}^{C},E_{V}^{R})$ 
 \\
 \hspace{1pt}$\epsilon_{Q}^{8}$\hspace{1pt} & $\frac{E_{V}^{R}}{4}-\frac{E_{V}^{L}}{4}$ & $-\frac{E_{V}^{C}}{2}+\frac{E_{V}^{L}}{4}+\frac{E_{V}^{R}}{4}$ & \hspace{4pt} max$(E_{V}^{L},E_{V}^{C},E_{V}^{R})$ \hspace{4pt}
 \\
 \hline\hline
\end{tabular}
\caption{Coordinates of the triple intersection points $\epsilon_{Q}^{n}$ of (1,0,0)-(0,1,0)-(0,0,1) charge states in detuning space $(\varepsilon,\varepsilon_{M})$ as a function of the valley splittings $E_{V}^{L}$, $E_{V}^{C}$, and $E_{V}^{R}$. (Forth column) Estimated excitation energy $E\st{ex}=E-E\st{GS}$ to populate the state with respect to the ground state energy at the triple points in the absence of tunneling. The triple points are only visible in the output signal if there is a finite population of the corresponding states. }
\label{4tab:intersection}
\end{center}
\end{table}

\subsubsection{Finite bias at low temperature}
Instead of the two-step measurement to extract the energy spectrum from cuts through the (1,0,0)-(0,1,0) and (0,1,0)-(0,0,1) charge configurations discussed above, an investigation of the charge intersection point (1,0,0)-(0,1,0)-(0,0,1) yields the same information about the valley splitting. This is especially interesting for measurements at low temperatures since the fine-structure of cuts through (1,0,0)-(0,1,0) and (0,1,0)-(0,0,1) charge transitions is invisible in the spectroscopic data due to the blocked charge current while current flow near the triple intersection points populates the necessary excited valley states (see appendix~\ref{4ssec:current}). Considering the same setup as above there are $n=2^{3}=8$ copies of the triple intersection point (1,0,0)-(0,1,0)-(0,0,1) due to the presence of the valley degree of freedom. The location of the intersection points $\epsilon_{Q}^{n}$ in detuning space $(\varepsilon,\varepsilon_{M})$ are shown in Table~\ref{4tab:intersection} together with an approximate energy necessary to populate the corresponding states. Each triple intersection point can be approximated for $(\varepsilon,\varepsilon_{M})=\epsilon_{Q}^{n}$ by the three-level system with eigenenergies
\begin{align}
E_{Q,1}^{n}&=\sqrt{|t_{x}^{n}|^{2}+|t_{y}^{n}|^{2}},\\
E_{Q,2}^{n}&=0,\\
E_{Q,3}^{n}&=-\sqrt{|t_{x}^{n}|^{2}+|t_{y}^{n}|^{2}},
\label{4eq:splittLeft}
\end{align}
where $t_{x}^{n}\in\lbrace t_{L},t_{L}^{\prime}\rbrace$ and $t_{y}^{n}\in\lbrace t_{R},t_{R}^{\prime}\rbrace$ depending on the intersection point. Close to these points the three-level system forms a coupled two-level system between the states $\ket{E_{Q,1}^{n}}$ and $\ket{E_{Q,3}^{n}}$ with the third state $\ket{E_{Q,2}^{n}}$ lying in the middle, where $\ket{E_{Q,i}^{n}}$ denotes the eigenstate with eigenenergy $E_{Q,i}^{n}$. The three-level system posses a large electric quadrupole between the eigenstates $\ket{E_{Q,1}^{n}}$ and $\ket{E_{Q,3}^{n}}$~\cite{Friesen2017,Koski2019}; all dipole moments are suppressed due to symmetry. Therefore, a symmetric architecture of the TQD resonator system, i.e, connecting the resonator via the center gate, is advantageous for probing these triple points. The probe frequency is ideally set to $\hbar \widetilde{\omega}_{P}=\hbar \widetilde{\omega}\st{res}=E_{Q,1}^{n}-E_{Q,3}^{n}\approx \sqrt{2} \omega_{P}$. 

Fig.~\ref{4fig:main3} shows the phase response of the probe signal in the vicinity of triple points for (a) $\Delta\mathcal{V}=\unit[0.3]{mV}$ and (b) $\Delta\mathcal{V}=\unit[-0.3]{mV}$. For an extraction of all three valley splittings a minimum of three triple points are necessary. If not enough features are visible in the phase response reversing the direction of the charge current helps to locate the position of missing triple intersection points. We see clearly a response in the phase at $\epsilon_{Q}^{1}=\unit[(-12,0)]{\mu eV}$, $\epsilon_{Q}^{2}=\unit[(-51,41)]{\mu eV}$, $\epsilon_{Q}^{4}=\unit[(49,61)]{\mu eV}$, and $\epsilon_{Q}^{7}=\unit[(49,-39)]{\mu eV}$. 
In total we extract with this method the valley splittings $E_{V}^{L}=\unit[82.4]{\mu eV}$, $E_{V}^{C}=\unit[100]{\mu eV}$, and $E_{V}^{R}=\unit[122.2]{\mu eV}$ which are roughly $3\%$ larger than the input settings $\widetilde{E}_{V}^{L}=\unit[80]{\mu eV}$, $\widetilde{E}_{V}^{C}=\unit[100]{\mu eV}$, and $\widetilde{E}_{V}^{R}=\unit[120]{\mu eV}$. We again attribute this error to a deformation of the energy levels due to the mixing of the different levels via tunneling and the broadening of the response signal. 

\begin{center}
\begin{figure}
\centering\igr[width=0.8\textwidth]{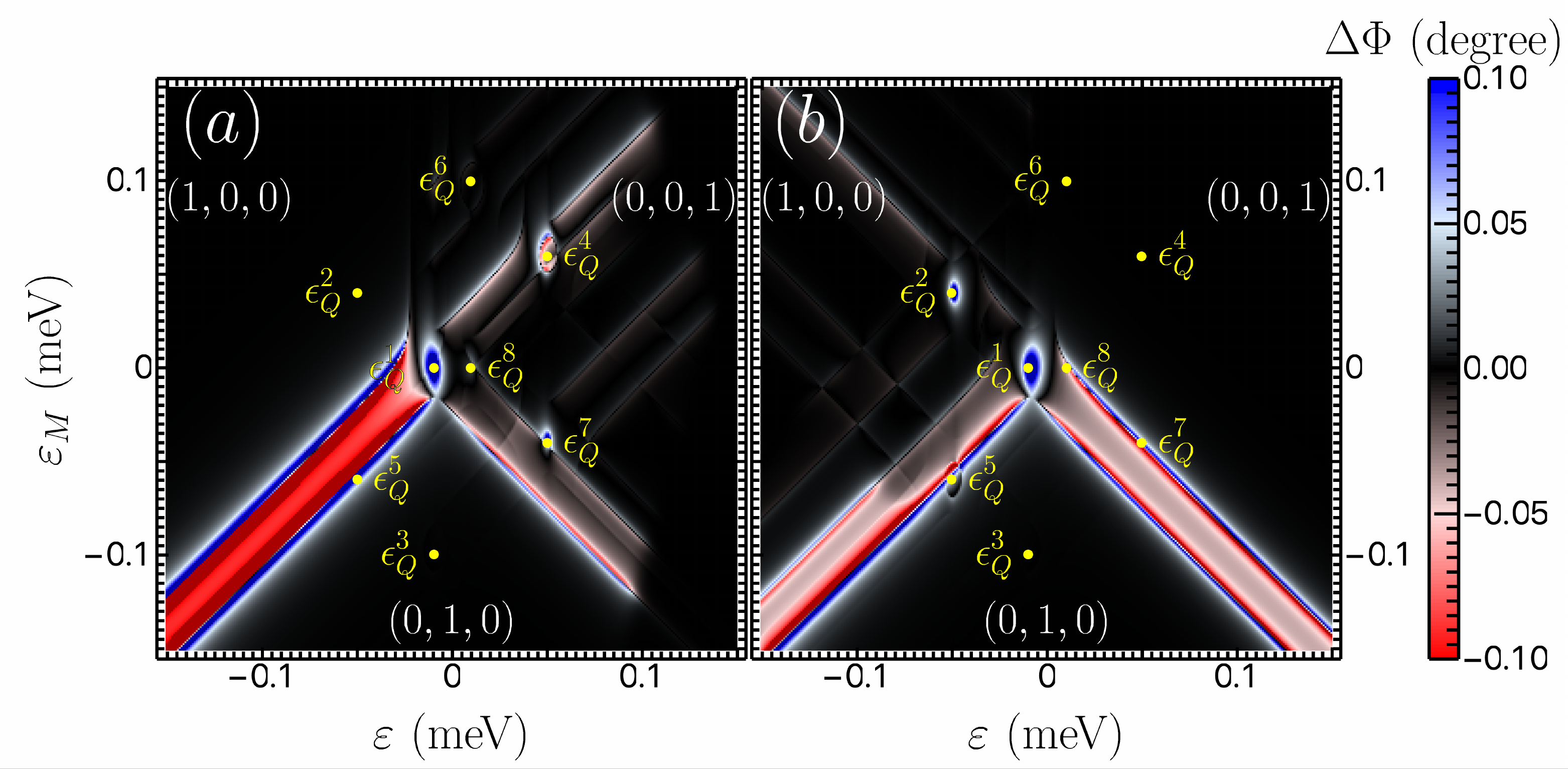}
\caption{Calculated phase response $\Delta\Phi=\text{arg}(A)$ of the probe signal of the TQD filled with a single electron coupled to the microwave cavity for $T=\unit[30]{mK}$ and applied voltage bias (a) $\Delta\mathcal{V}=\unit[0.3]{mV}$ and (b) $\Delta\mathcal{V}=\unit[-0.3]{mV}$ as a function of the TQD detuning parameters $\varepsilon$ and $\varepsilon_{M}$. The resonator and the probe frequency $\widetilde{\omega}_{P}=\sqrt{2}\omega_{P}$ and $\widetilde{\omega}\st{res}=\sqrt{2}\omega\st{res}$ are adjusted to probe the charge quadrupole transitions~\cite{Friesen2017,Koski2019} at the triple intersection points $\epsilon_{Q}^{i}$ (yellow). The positions of $\epsilon_{Q}^{1}$, $\epsilon_{Q}^{2}$, $\epsilon_{Q}^{4}$, and $\epsilon_{Q}^{7}$ (see Table~\ref{4tab:intersection}) are sufficient to extract the values of all three valley splittings $E_{V}^{L}$, $E_{V}^{C}$, and $E_{V}^{R}$.}
  \label{4fig:main3}
\end{figure}
\end{center}

\subsection{Three electrons in a triple quantum dot}
\label{4ssec:three}

In practice studying the three-electron case is more interesting since it allows one to measure the valley splitting and valley phase in the same charge configuration regime spin qubits can be implemented, i.e., three spin-$\frac{1}{2}$ qubits or a exchange-only qubit are implemented in the (1,1,1) charge regime.
The total Hamiltonian of the three-electron case is given by 
\begin{align}
\h_{3e}=H\st{charge,3e}+H\st{valley,3e}+H\st{Zeeman,3e}+H\st{tunnel,3e}
\label{4eq:Ham3e}
\end{align}
which can be obtained from the Hamiltonian~\eqref{4eq:HamTot} with $N=3$. Focusing only on the (2,0,1), (1,1,1), and (1,0,2) charge configuration regime where the resonant exchange (RX) qubit is typically realized, the charge Hamiltonian can be simplified to
\begin{align}
H\st{charge,3e}=\left(
\begin{matrix}
\varepsilon-\varepsilon_{M}& 0 & 0\\
0 & 0& 0\\
0& 0 &-\varepsilon-\varepsilon_{M}
\end{matrix}\right)+3\varepsilon_{g}\mathds{1}_{3}
\end{align}
containing the electrostatic interactions from the capacitor model. The detuning parameters $\varepsilon$ and $\varepsilon_{M}$ are (up to a constant energy shift) identical to Eqs.~\eqref{4eq:det1ed}~and~\eqref{4eq:det1em}.
We choose the average detuning $\varepsilon_{g}$ such that the TQD is occupied by three electrons. The left-center and right-center detuning parameters, $\varepsilon_{L}$ and $\varepsilon_{R}$, then allow us to investigate the (2,0,1)-(1,1,1) and (1,0,2)-(1,1,1) charge transitions. Analogously to the single-electron case, the dynamics is effectively reduced to an DQD filled with two electrons. 

\begin{center}
\begin{figure}
\centering\igr[width=0.8\textwidth]{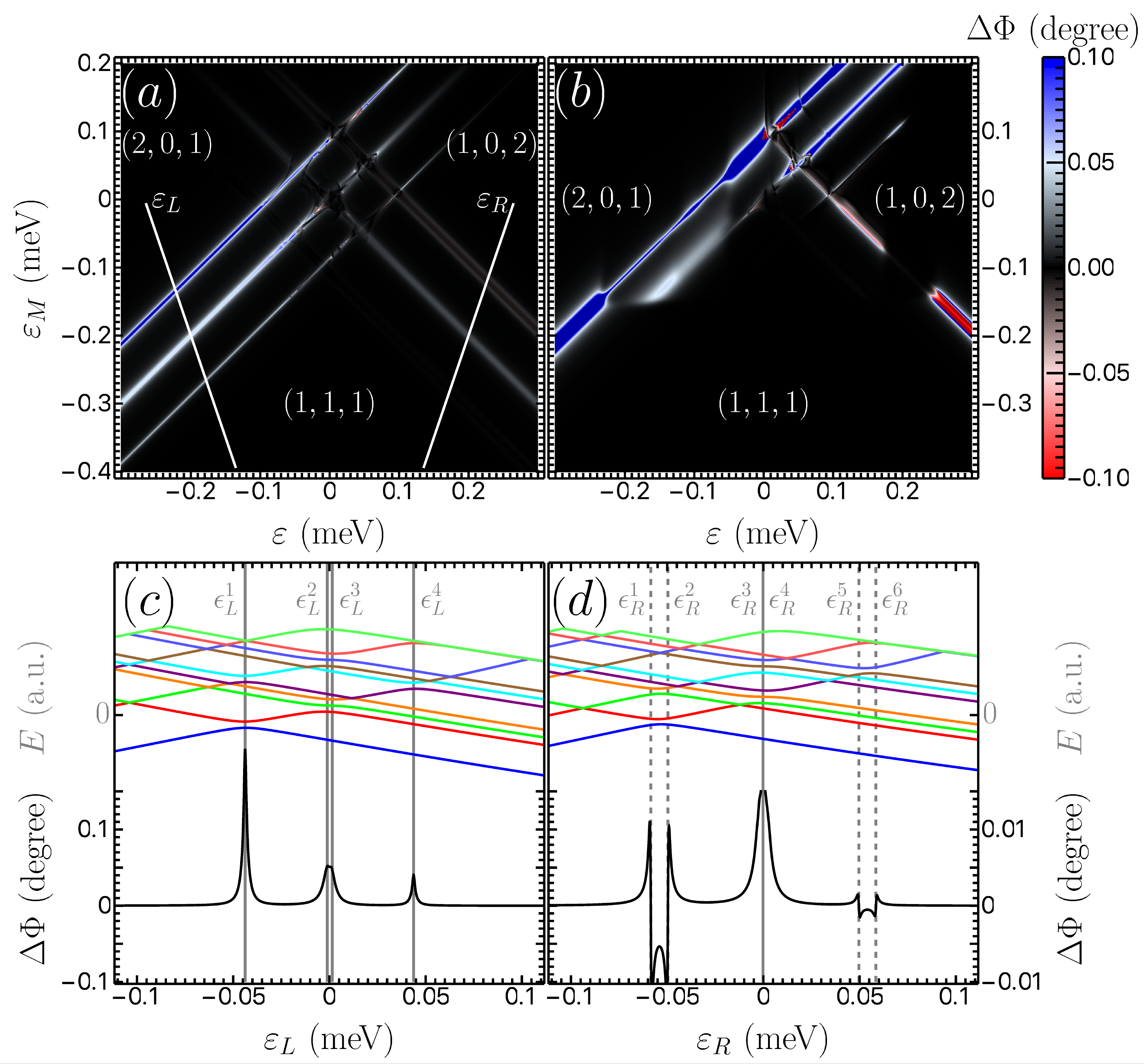}
\caption{(a) Calculated phase response $\Delta\Phi=\text{arg}(A)$ of the probe signal of the TQD coupled to the microwave cavity for $T=\unit[1]{K}$ and no applied voltage bias $\Delta\mathcal{V}=0$. Here $(n_{L},n_{C},n_{R})$ denotes the occupation of dot $i=L,C,R$ with $n_{i}$ with electrons. The left and right white lines are cuts along the double quantum dot (DQD) detuning parameters $\varepsilon_{L}$ and $\varepsilon_{R}$ while keeping $\varepsilon_{g}$ and the respective orthogonal detuning parameter fixed. (b) Calculated phase response for $T=\unit[30]{mK}$ and applied bias $\Delta\mathcal{V}=\unit[0.3]{mV}$.  This allows for the investigation of the (2,0,1)-(1,1,1) and (1,0,2)-(1,1,1) charge transition.
 (c) Cut along $\varepsilon_{L}$ [white line in (a)] and the energy of the ten lowest eigenstates, $E_{i}$, is plotted as function of $\varepsilon_{L}$. 
 (d) Cut along $\varepsilon_{R}$ [white line in (a)]. For interpretation also the energy of the ten lowest eigenstates are plotted as function of $\varepsilon_{R}$. The sharp peaks $\epsilon_{R}^{1}$ and $\epsilon_{R}^{2}$ correspond to the condition $E_{2}-E_{1}=\hbar\omega\st{res}$.}
  \label{4fig:main4}
\end{figure}
\end{center}

\subsubsection{Extracting the valley splitting and phase}

The valley degeneracy effectively creates eight copies of the charge states, two from the valley DOF in each dot for the (1,1,1) configuration and two copies for the (2,0,1) and (1,0,2) configuration neglecting the spin. These states are coupled by the valley non-conserving tunnel matrix elements. Therefore, instead of a single anti-crossing between charge states we expect to see (in the ideal case) 16 anti-crossings between the (2,0,1)-(1,1,1) and (1,0,2)-(1,1,1) charge states. Of course, to observe all crossings requires a temperature or bias such that the excited states are populated. In Fig.~\ref{4fig:main4}~(a) and~(b) the phase shift of the cavity signal for three electrons is shown as a function of the two detuning parameters $\varepsilon,\varepsilon_{M}$. At the (2,0,1)-(1,1,1) and the (1,0,2)-(1,1,1) charge transitions we could potentially see the splitting of a single line into multiple lines. The asymmetry in brightness between the (2,0,1)-(1,1,1) and the (1,0,2)-(1,1,1) charge transitions is due to different energy detunings $\Delta=|(E_{2}-E_{1})-\hbar\omega\st{res}|$ for the left and right charge transitions. 

A cut along the left-center detuning $\varepsilon_{L}$ provides information about the level splittings [see Fig.~\ref{4fig:main4}~(c)]. The ground state in the (1,1,1) regime is a polarized valley state, where all electrons occupy the lower valley state, and in the (2,0,1) charge regime the two electrons form a valley-singlet state and the remaining electron in the right dot occupies the ground state. The respective energy level crossing occurs at $\varepsilon_{L}=-(E_{V}^{L}+E_{V}^{C})/4$. From Fig.~\ref{4fig:main4}~(c) we find $\epsilon_{L}^{1}=-(E_{V}^{L}+E_{V}^{C})/4$, $\epsilon_{L}^{2}\approx\epsilon_{L}^{3}\approx0$, and $\epsilon_{L}^{1}=(E_{V}^{L}+E_{V}^{C})/4$ which are all consistent with the extracted valley splitting in the single electron case.

A cut along the right-center detuning $\varepsilon_{R}$ between (1,0,2) and (1,1,1) charge states shows similar features [see Fig.~\ref{4fig:main4}~(d)]. The respective energy crossing occurs at $\varepsilon_{R}=-(E_{V}^{R}+E_{V}^{C})/4$ and we find again two surrounding peaks at $\epsilon_{R}^{1}$ and $\epsilon_{R}^{2}$. From Fig.~\ref{4fig:main4}~(d) we find $-(E_{V}^{R}+E_{V}^{C})/4=(\epsilon_{R}^{1}+\epsilon_{R}^{2})/2$, $\epsilon_{R}^{3}\approx\epsilon_{R}^{4}\approx0$, and $(E_{V}^{R}+E_{V}^{C})/4=(\epsilon_{R}^{5}+\epsilon_{R}^{6})/2$. This matches with the results in the single electron case.

Unfortunately, further energy crossings are hardly visible for our choice of simulation parameters in the case of three electrons, thus, we refrain from a further analysis. 

\section{Conclusion}
\label{4sec:conclusion}

In this paper we have theoretically investigated the response signal of a probed microwave resonator coupled to a linearly arranged triple quantum dot via the center dot gate electrode. A realistic model of the TQD is used in our analysis which includes electrostatic cross-talk between the dots and gates via a capacitor model, valley and spin effects, and the solution of a Redfield master equation to find the occupation probabilities. We show that a setup consisting of a TQD filled with a single electron can be used to extract important information from the TQD system such as the valley splitting and the valley phase. The accuracy of the extracted valley splitting and phase becomes higher and the interpretation simpler if the TQD is detuned such that one chemical potential is significantly increased which reduces the triple dot system to an effective double dot. A setup consisting of three electrons in a triple quantum dot is in principle capable to deliver the same information but the larger number of energy levels makes the population of the relevant excited valley states and the corresponding interpretation of the signal more difficult.

\ack
We acknowledge funding from ARO through Grant No. W911NF-15-1-0149 and the DFG through SFB 767. We thank M. Benito and F. Ginzel for helpful discussions. We thank J. Petta for providing us with experimental datasets. 


\appendix


\section{Secular approximation}

In order to compute the occupation of the energy levels we relied on the secular approximation. However, since we have no all-to-all coupling there are energy levels which do not form an anti-crossing. At these points the energy splitting goes to zero, $|E_{i}-E_{j}|\rightarrow 0$, thus, violating the secular approximation. The validity of our calculation, however, is unaffected since the ratio between the number of valid points $N_{G}$ and detected violations $N_{F}$ is small for large sample sizes $N_{S}$, $N_{G}/N_{F}\gg 1$. In all simulation we have used $N_{S}=600^{2}$ sample points. 

\section{Charge current}
\label{4ssec:current}
As discussed in Ref.~\cite{Burkard2016}, excited energy states required for read-out of all relevant system parameters can be populated either by increasing the temperature in the system or by applying a dc bias voltage. While precise control over the temperature is experimentally challenging, biasing the left and right leads is not.
The charge current from left to right can be given in two equivalent forms due to continuity
\begin{subequations}
\bean
I&=&e\Gl \sum_{m\neq n} \left(c_{mLn} - c_{nLm}\right)n_{nLm} \overline{\rho}_m \\
&=&e\Gr \sum_{m\neq n} \left(c_{nLm} - c_{mLn} \right)n_{nLm} \overline{\rho}_m,
\eean
\end{subequations}
where summations run for all $m$ and $n$. The expressions for a charge current from right to left is similar. Fig.~\ref{4fig:supp1} shows the charge current for $\Delta\mathcal{V}=\unit[\pm 0.3]{mV}$. A finite current is only possible at charge quadruple points~\cite{Schroeer2007} where four charge configurations intersect which in our case is in the vicinity of the triple intersection points $\epsilon_{Q}^{n}$.

\begin{center}
\begin{figure}
\centering\igr[width=0.8\columnwidth]{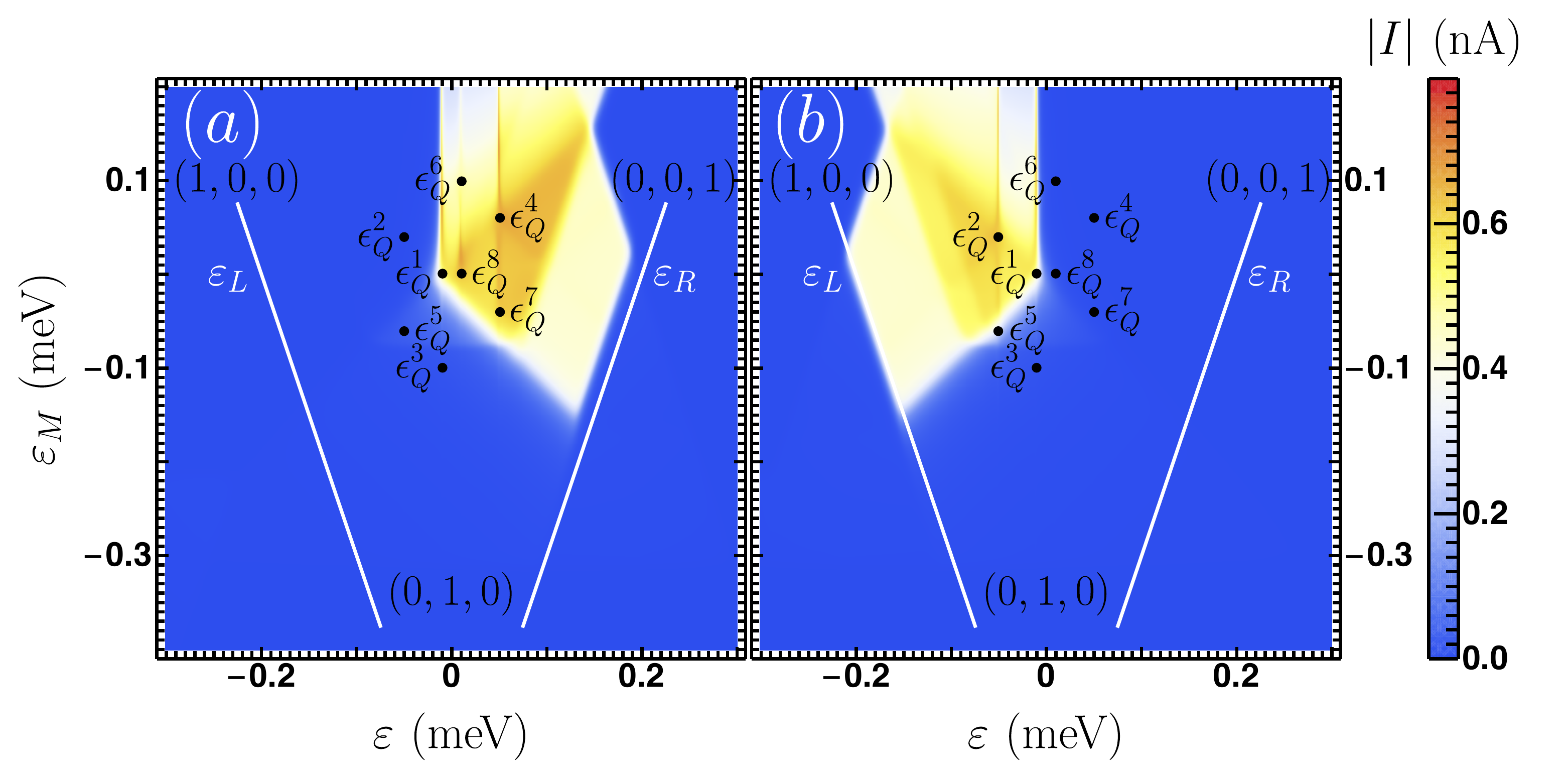}
\caption{Calculated charge current $I$ of the TQD coupled to the microwave cavity for $T=\unit[30]{mK}$ and applied voltage bias (a) $\Delta\mathcal{V}=\unit[0.3]{mV}$ and (b) $\Delta\mathcal{V}=\unit[-0.3]{mV}$. Here $(n_{L},n_{C},n_{R})$ denotes the occupation of dot $i=L,C,R$ with $n_{i}$ with electrons. The left and right white lines are cuts along the double quantum dot (DQD) detuning parameters $\varepsilon_{L}$ and $\varepsilon_{R}$ while keeping $\varepsilon_{g}$ and the respective orthogonal detuning parameter fixed. The black dots mark the triple intersection points $\epsilon_{Q}^{i}$ (black). Note, that in (b) the charge direction is reversed.}
  \label{4fig:supp1}
\end{figure}
\end{center}

\section{Simulation parameters}
\label{4ch:app}
For the simulation in the main text we use the following parameters from experiments in undoped Si/SiGe performed in a triple quantum dot using the gate layout described in Ref.~\cite{Zajac2015}. The extracted capacitance matrix consisting of the electrostatic capacitances between the dots reads 
\begin{align}
\mathcal{C}_{\text{dot}}=\left(
\begin{array}{ccc}
 56.2 & -5.5 & -0.5 \\
 -5.5 & 50.5 & -11.7 \\
 -0.5 & -11.7 & 59.4 \\
\end{array}
\right)
\end{align}
and the extracted capacitance matrix consisting of the electrostatic capacitances between the dots and the gates reads 
\begin{align}
\mathcal{C}_{\text{gate}}=\left(
\begin{array}{ccc}
 -6.9 & -2.4 & -0.3 \\
 -0.15 & -5.9 & -0.03 \\
 -0.4 & -3.6 & -6.9 \\
\end{array}
\right).
\end{align}
The capacitance matrix consisting of the electrostatic capacitances between the dots and the leads is set to
\begin{align}
\mathcal{C}_{\text{lead}}=\left(
\begin{array}{cc}
 40.6 & 0 \\
 13.6 & 13.6 \\
 0 & 36.4 \\
\end{array}
\right).
\end{align}
All capacitances are given in units of (aF) attofarad. 

The remaining parameters for the simulation are the tunneling couplings, $t_{12}$ and $t_{23}$, between the dots, the valley-orbit parameters $\Delta_{j}=E_{V}^{j}\,\E^{i\phi_{j}}$ in each dot $j=L,C,R$, the incoherent decay rates $\gamma_{c}$ and $\gamma_{v}$, and the charge dephasing rate $\gamma\st{dep}$. The tunneling parameters used in all simulations in the main text are chosen to be $t_{12}=\unit[12.5]{\mu eV}$ and $t_{23}=\unit[11.5]{\mu eV}$. For the valleys splittings we use $E_{V}^{L}=\unit[80]{\mu eV}$, $E_{V}^{C}=\unit[100]{\mu eV}$, and $E_{V}^{R}=\unit[120]{\mu eV}$. The relative valley phases $\theta_{LC}=\phi_{L}-\phi_{C}=\unit[0.23]{\pi}$, $\theta_{RC}=\phi_{R}-\phi_{C}=\unit[-0.2]{\pi}$, and $\theta_{LR}=\phi_{L}-\phi_{R}=\unit[0]{\pi}$, where the last phase is undetectable in a linear aligned triple quantum dot.
The decay and dephasing rates are set to $\gamma_{c}=\unit[0.12]{\mu eV}$, $\gamma_{c}=\unit[0.12]{\mu eV}$, and $\gamma\st{dep}=\unit[1.2]{\mu eV}$.

\clearpage
\bibliographystyle{iopart-num}
\bibliography{Lit_07_19}

\end{document}